\documentclass[twocolumn,showpacs,preprintnumbers,amsmath,amssymb]{revtex4}

\usepackage{graphicx}
\usepackage{dcolumn}
\usepackage{bm}
\newcommand{\sgn}{\mathop{\mathrm{sgn}}}

\begin{document}

\preprint{CHSH micr}

\title{A probability loophole in the CHSH.}

\author{Han Geurdes}
\affiliation{C. vd Lijnstraat 164, 2593 NN Den Haag Netherlands}


\date{\today}

\begin{abstract}
In the present paper a robustness stress-test of the CHSH experiments for Einstein locality and causality is designed and employed. Random A and B from dice and coins, but based on a local model, run "parallel" to a real experiment. We found a local causal model with a nonzero probability to violate the CHSH inequality for some relevant quartets $\mathcal{Q}$ of settings in the series of trials. 
\end{abstract}

\pacs{03.65.Ud, 03.67.Lx, 03.67.Dd}
\maketitle
\section{Introduction \& model}
The statistical basis of CHSH conclusion is studied. In 1964 J.S. Bell postulated the local hidden variables (LHV) correlation $E(a,b)=\int_{\lambda \in \Lambda} \rho_{\lambda}A_{\lambda}(a)B_{\lambda}(b)d \lambda$. Here, $A_{\lambda}(a)$ and $B_{\lambda}(b)\in \{-1,1\}$. For more details see \cite{Bell}. Clauser \cite{Claus69} derived the CHSH inequality $|S|\leq 2$ thereof with
\begin{equation}
S=E(1_A,1_B)-E(1_A,2_B)-E(2_A,1_B)-E(2_A,2_B).
\end{equation}
In the CHSH, setting pairs $\mathcal{Q}=\mathcal{A}\times \mathcal{B}$ are used with $a \in \{1_A,2_A\}=\mathcal{A}$ and $b \in \{1_B,2_B\}=\mathcal{B}$. The $|S|\leq 2$ \underline{must be} valid for all LHV models for each trial, at any moment. Our settings are the violating pairs $1_A=(1,0,0)$, $2_A=(0,1,0)$, $1_B=\frac{1}{\sqrt{2}}(1,-1,0)$ and $2_B=\frac{1}{\sqrt{2}}(-1,-1,0)$. 

If $|S|>2$ is found in nature such as in \cite{Weihs}, then under local conditions, Einstein locality \cite{EPR} does not occur. Of course, this is valid only if $|S|>2$ is LHV impossible. We reformulate this as $\Pr\{|S|>2\,|\,using \,LHV\}=0$. This must be true for each relevant $\mathcal{Q}$ in each experiment for each model in order for a CHSH experiment to make sense. 

In the proposed stress-test, Alice and Bob {\it{simulate}} a "parallel" independent $A$ and $B$ sequence with additional coins and dices \cite{Geur}. Let us define $3$ sets: $\Omega_{+}(a,b,x,y):= \{\lambda \in \Lambda | A_{\lambda}(a)B_{\lambda}(b)=A_{\lambda}(x)B_{\lambda}(y)=+1\},\,
\Omega_{-}(a,b,x,y):=\{\lambda \in \Lambda | A_{\lambda}(a)B_{\lambda}(b)=A_{\lambda}(x)B_{\lambda}(y)=-1\}
$
and
$
\Omega_{0}(a,b,x,y):=\{\lambda \in \Lambda | A_{\lambda}(a)B_{\lambda}(b)=-A_{\lambda}(x)B_{\lambda}(y)=\pm 1\}
$
with, $\Lambda=\Omega_0 \cup \Omega_{+} \cup \Omega_{-}$ and the $\Omega$ sets disjoint. Hence, $E(a,b)-E(x,y)=\int_{\lambda \in \Lambda}\{A_{\lambda}(a)B_{\lambda}(b)-A_{\lambda}(x)B_{\lambda}(y)\}\rho_{\lambda}d \lambda$ only $\lambda \in \Omega_{0}$ do not cancel. $\Rightarrow E(a,b)-E(x,y)=-2\int_{\lambda \in \Omega_{0}(a,b,x,y)}A_{\lambda}(x)B_{\lambda}(y)\rho_{\lambda}d \lambda$. Suppose, $a,b \notin \mathcal{A} \cup \mathcal{B}$ and $(a,b)$ such that $E(a,b)=0$ then  
\begin{equation}
\frac{E(x,y)}{2}=\int_{\lambda \in \Omega_{0}(a,b,x,y)}A_{\lambda}(x)B_{\lambda}(y)\rho_{\lambda}d \lambda.
\end{equation}
This is $E_{T}(x,y)$. Because $E(a,b)=0$, consistency requires 
\begin{equation}
\frac{E(x,y)}{2}=\int_{\lambda \in \Omega_{+}(a,b,x,y)}\rho_{\lambda}d \lambda-\int_{\lambda \in \Omega_{-}(a,b,x,y)}\rho_{\lambda}d \lambda.
\end{equation}
This is $E_C(x,y)$. Numerically: $E_{T}(x,y)\approx E_{C}(x,y)$. 

Suppose, $\rho_{\lambda}=\rho_{\lambda_1}\rho_{\lambda_2}$ and $\lambda_1$, is assigned to Alice's measuring instrument, $\lambda_2$ to Bob's. $\rho_{\lambda_j}=\frac{1}{\sqrt{2}}$ for $\lambda_j \in \Lambda_j$, $\Lambda_j=\{\lambda_j | \frac{-1}{\sqrt{2}} \leq \lambda_j \leq \frac{1}{\sqrt{2}} \}$ and zero "elsewhere" $(j=1,2)$; $\Lambda =\Lambda_1 \times \Lambda_2$. The $E_T$ and $E_C$ transform in 
\begin{equation}
E_T(x,y)=\int_{(\lambda_1,\lambda_2)\in \Omega_0(a,b,x,y)}A_{\lambda_1}(x)B_{\lambda_2}(y)
\end{equation}
and 
\begin{equation}
E_C(x,y)=\int_{(\lambda_1,\lambda_2)\in \Omega_{+}(a,b,x,y)}-\int_{(\lambda_1,\lambda_2)\in \Omega_{-}(a,b,x,y)}
\end{equation}
$\int$ represents double integration where necessary. Further, where possible, $d\lambda_1 d \lambda_2$ is suppressed. The $A$ function, for $x \in \mathcal{A}$ is given by
\begin{eqnarray}\label{e12}
A_{\lambda_1}(x)=\{
\begin{array}{ll}
\alpha_{\lambda_1}(x),~~\lambda_1 \in I(x)\\
\sgn[\zeta(x)-\lambda_1],~\lambda_1 \in (\Lambda_1 \backslash I(x))
\end{array}
\end{eqnarray}
Here, $I(1_A)=\{\lambda_1|-\frac{1}{\sqrt{2}}\leq \lambda_1 \leq 1- \frac{1}{\sqrt{2}}\}$, $I(2_A)=\{\lambda_1|-1+\frac{1}{\sqrt{2}}\leq \lambda_1 \leq  \frac{1}{\sqrt{2}}\}$. For $y \in \mathcal{B}$,
\begin{eqnarray}\label{e13}
B_{\lambda_2}(y)=\{
\begin{array}{ll}
\beta_{\lambda_2}(y),~~\lambda_2 \in J(y)\\
\sgn[\eta(y)-\lambda_2],~\lambda_2 \in (\Lambda_2 \backslash J(y))
\end{array}
\end{eqnarray}
$J(1_B)=\{\lambda_2|-\frac{1}{\sqrt{2}}\leq \lambda_2 \leq 0\}$, $J(2_B)=\Lambda_2\backslash J(1_B)$, $\sgn(0)=1$ and $\sgn(x)=\frac{x}{|x|}, (x\neq 0)$. 

Suppose Alice tossed $1_A$ and Bob $1_B$. The $\alpha$ and $\beta$ are determined by tossing a fair coin. Heads is $+1$ and tails is $-1$. Hence, $\Pr_{coins}\{\alpha_{\lambda_1}(1_A)\beta_{\lambda_2}(1_B)=-1\}>0$. Moreover, $\forall\,(x,y)\in \mathcal{Q}\backslash \{(1_A,1_B)\}$ and $\Pr_{coins}\{\alpha_{\lambda_1}(x)\beta_{\lambda_2}(y)=1 \}>0$. In  addition to the coins Alice and Bob each hold a $4$-sided dice to determine the $V$ and $U$ functions  viz. $\zeta$ and $\eta$ in (\ref{e12}) and (\ref{e13}). Suppose that Carrol, by a draw from the model-pool, determines the employed model. She sees $\Pr_{pool}\{\Omega_{+}(a,b,1_A,1_B)=\emptyset \,\&\, \Omega_{-}(a,b,1_A,1_B)=I(1_A)\times J(1_B)\}>0$. Hence, $\Pr_{E-space}\{E_C(1_A,1_B)=\frac{-1}{\sqrt{2}}\}>0$. E-space is the combination of pool, coins and dices probability spaces. For $E_T$ let us look at $\Pr_{pool}\{ \Omega_{0}(a,b,1_A,1_B)=((\Lambda_1\backslash I(1_A))\times J(1_B))\cup ((\Lambda_1\backslash I(1_A))\times J(2_B))\cup (I(1_A)\times J(2_B))\} >0$.
Note, $\Omega_0\cup\Omega_{+}\cup\Omega_{-}=\Lambda$. It follows that  
\begin{widetext}
\begin{equation}
E_T(1_A,1_B)=\int_{(\lambda_1,\lambda_2)\in I(1_A)\times J(2_B)}\alpha u(\lambda_2)+\int_{(\lambda_1,\lambda_2) \in (\Lambda_1 \backslash I(1_A))\times J(1_B)}\beta v(\lambda_1)+\int_{(\lambda_1,\lambda_2)\in (\Lambda_1\backslash I(1_A))\times J(2_B)}v(\lambda_1)u(\lambda_2).
\end{equation}
\end{widetext}
Here, $u(\lambda_2)=\sgn[\eta(1_B)-\lambda_2]\}$ and $v(\lambda_1)=\sgn[\zeta(1_A)-\lambda_1]$. 
Note, $\int_{I(1_A)}d\lambda_1=\int_{I(2_A)}d\lambda_1=1$ and $\int_{J(1_B)}d\lambda_2=\int_{J(2_B)}d\lambda_2=\frac{1}{2}\sqrt{2}$. The more general expression $E_T(x,y)=\alpha U(y)+ \frac{\beta}{\sqrt{2}}V(x) +U(y)V(x)$ can subsequently be derived from the previous $uv$ equation. Note, $U(y)=\int_{(\Lambda_2\backslash J(y))}\sgn[\eta(y)-\lambda_2]d\lambda_2$ and $V(x)=\int_{(\Lambda_1\backslash I(x))}\sgn[\zeta(x)-\lambda_1]d\lambda_1$. For $V(x)$, $V(1_A)=2\zeta(1_A)-1$ and $V(2_A)=2\zeta(2_A)+1$. Because of its use as {\it random} function, $V(x) \in [1-\sqrt{2}, \sqrt{2}-1]\approx (-0.4142,0.4142)$. And, $U(1_B)=2\eta(1_B)-\frac{1}{\sqrt{2}}$ and $U(2_B)=2\eta(2_B)+\frac{1}{\sqrt{2}}$ with $U(y)\in \Lambda_2 \approx (-0.7071,0.7071)$. 

The random functions $\zeta$ and $\eta$ translate to 4-sided dices for the sign integrals $V$ and $U$. We determine the numerical values for $U$ and $V$ below. Suppose $(1_A,1_B)$ then the $E$ thereof can be rewritten as ($\alpha\beta=-1$) $U(1_B)-\frac{1}{\sqrt{2}} V(1_A)+\alpha U(1_B)V(1_A)=-\frac{\alpha}{\sqrt{2}}$. For $\alpha=1,\beta=-1$ computation gave $U(1_B)\approx -0.45371$ and $V(1_A)\approx 0.218186$ with error $\delta(U,V)=|U-(V/\sqrt{2}) +\alpha UV - \alpha E_{QM}(a,b)| \approx 9.9\times 10^{-7}$. Here $E_{QM}=\frac{-1}{\sqrt{2}}$. For $\alpha=-1, \beta=1$, $U(1_B)\approx 0.32760$ and $V(1_A)\approx -0.36691$ with error $\delta(U,V)=8.0\times 10^{-7}$. Hence, an approximate consistency in probability: $\Pr_{E-space}\{E_T(1_A,1_B)\approx \frac{-1}{\sqrt{2}}\}>0$. This leads to, $\Pr_{E-space}\{E(1_A,1_B)= \frac{-1}{\sqrt{2}}\}>0$. If Alice tosses $1_A$ and Bob $2_B$ then when Alice tosses her $\alpha$ and Bob his $\beta$; $\Pr_{coins}\{\alpha_{\lambda_1}(1_A)\beta_{\lambda_2}(2_B)=1 \}>0$. Carrol draws from the LHV model pool and has $\Pr_{pool}\{\Omega_{+}(a,b,1_A,2_B)=I(1_A)\times J(2_B) \,\&\,\Omega_{-}(a,b,1_A,2_B)=\emptyset \}>0$. We arrive at $\Pr_{E-space}\{E_C(1_A,2_B)=\frac{1}{\sqrt{2}}\}>0$. In order to determine $E_T(1_A,2_B)$ the $\Omega_0$ shows $\Pr_{pool}\{ \Omega_{0}(a,b,1_A,2_B)=((\Lambda_1\backslash I(1_A))\times J(1_B))\cup ((\Lambda_1\backslash I(1_A))\times J(2_B))\cup (I(1_A)\times J(1_B))\} >0$. For $(1_A,2_B)$, $E$ is ($\alpha\beta=1$): $U(2_B)+\frac{1}{\sqrt{2}}V(1_A)+\alpha U(2_B)V(1_A)=\frac{\alpha}{\sqrt{2}}$. For $\alpha=1, \beta=1$ we have $U(2_B)\approx 0.3001$ and $V(1_A)\approx 0.4042$ with $\delta(U,V)\approx 3.4 \times 10^{-5}$. Here, $E_{QM}=\frac{1}{\sqrt{2}}$. For $\alpha=-1, \beta=-1$ we found $U(2_B)\approx -0.67710$ and $V(1_A)\approx -0.0216$ and $\delta(U,V)\approx 8.0 \times 10^{-7}$. Hence, $\Pr_{E-space}\{E_T(1_A,2_B)\approx \frac{1}{\sqrt{2}}\}>0$, or, $\Pr_{E-space}\{E(1_A,2_B)= \frac{-1}{\sqrt{2}}\}>0$. Note that for $(2_A, 1_B)$ and $(2_A,2_B)$ a similar form for $E(x,y)$ obtains as for $(1_A,2_B)$. The stress-test amounts to: Alice and Bob determine the setting $(x,y) \in \mathcal{Q}$  and record their spin. At any moment, Alice and Bob may toss $\alpha$ and $\beta$ coins and throw the dices; $V=(0.218186,-0.36691,0.4042,-0.0216)$ for Alice and $U=(-0.45371,0.32760,0.3001,-0.6771)$ for Bob and make a record using a trial number. Similarly for Carrol's draws from the model-pool. 
\section{Conclusion}
For $k=1,2,3,4, \exists_{n_k\in\{1,...,N\}}\exists_{(x,y)_{n_k}\in\mathcal{Q}} Pr\{E_{T(C)}(x,y)_{n_k}=E_{QM}(x,y)_{n_k}|\, LHV \}>0$, and, $(x,y)_n$ the n-th pair of settings $x$ and $y$. Hence,
\begin{eqnarray}\begin{array}{ll}\label{Prob}
\Pr \{|S|>2\,|\,LHV\}=\prod_{k=1}^4\prod_{(x,y)_{n_k}\in\mathcal{Q}; \,n_k\in \{1,..N\}} [ \\
~~~~~~\Pr\{E_{T(C)}(x,y)_{n_k}=E_{QM}(x,y)_{n_k}|LHV \}\,] >0
\end{array}
\end{eqnarray}
and $\,n_1\neq n_2, n_2 \neq n_3, n_3 \neq n_4, n_4\neq n_1$.  {The urn,} 
 {four sided dices and the $\alpha \,\&\, \beta$ coins connect   with non-}
 {zero probability the LHV elements of the model. The }
 {probability in (\ref{Prob}) is based on {\it per trial} probabilities. In} 
 {addition, the complete model, i.e. stress test plus LHV}    {part, is Kolmogorovian.} Hence, a probability loophole in the CHSH is found.  
 {The methodology presented is} 
 {valid for all settings.} The stress-test can be performed at any time. In e.g. experiment \cite{Weihs} , a nonzero probability exists that the violation {\it per quartet} of settings is obtained with LHV.  {The aim of the experiment is} 
 {the explanation of the (per quartet) entanglement.}
 {Hence, because of (\ref{Prob}) the CHSH no-go for LHV is flawed.} 
 {One can always point at trial numbers where nature} 
 {{\it could} have used LHV for $|S|>2$, because of the result} 
 {in equation (\ref{Prob}). The breach found in CHSH cannot} 
 {be plugged using LHV impossibilities construed with} 
 {CHSH principles. Moreover, LHV impossibilities using other} 
 {means do not take away that LHV may occur in nature} 
 {in the CHSH type experiments.} If (\ref{Prob}) does not include LHV, then CHSH does not exclude LHV.

\end{document}